# Millimetre-scale magnetocardiography of living rats using a solid-state quantum sensor

Keigo Arai[1,2,*], Akihiro Kuwahata[3,4,*], Daisuke Nishitani[1,*], Ikuya Fujisaki[1], Ryoma Matsuki[1], Zhonghao Xin[3], Yuki Nishio[1], Xinyu Cao[3], Yuji Hatano[1], Shinobu Onoda[5], Chikara Shinei[6], Masashi Miyakawa[6], Takashi Taniguchi[7], Masatoshi Yamazaki[8,9], Tokuyuki Teraji[6], Takeshi Ohshima[5], Mutsuko Hatano[1,5,†], Masaki Sekino[3,†], and Takayuki Iwasaki[1,†]

[1] School of Engineering, Tokyo Institute of Technology, Tokyo, 152-8550, Japan
[2] Japan Science and Technology Agency, PRESTO, Saitama, 332-0012, Japan
[3] Graduate School of Engineering, The University of Tokyo, Tokyo, 113-8656, Japan
[4] Graduate School of Engineering, Tohoku University, Sendai, 980-8579, Japan
[5] National Institutes for Quantum and Radiological Science and Technology, Takasaki, 370-1292, Japan
[6] Research Center for Functional Materials, National Institute for Materials Science, Tsukuba, 305-0044, Japan
[7] International Center for Materials Nanoarchitectonics, National Institute for Materials Science, Tsukuba, 305-0044, Japan
[8] Department of Cardiology, Nagano Hospital, Soja, 719-1126, Japan
[9] Medical Device Development and Regulation Research Center, The University of Tokyo, Tokyo, 113-8656, Japan

* These authors contributed equally to this work.
† Correspondence to hatano.m.ab@m.titech.ac.jp, sekino@g.ecc.u-tokyo.ac.jp, and iwasaki.t.aj@m.titech.ac.jp

**A key challenge in cardiology is the non-invasive imaging of electric current propagation occurring in the cardiovascular system at an intra-cardiac scale. A promising approach for directly mapping the current dynamics is to monitor the associated stray magnetic field. However, in this magnetic field approach, the spatial resolution deteriorates significantly as the standoff distance between the target and the sensor increases. Existing sensors usually remain relatively far from the target and provide only centimetre-scale resolution because their operating temperature is not biocompatible. Here we demonstrate millimetre-scale magnetocardiography of living rats using a solid-state quantum sensor based on nitrogen-vacancy centres in diamond. The essence of the method is a millimetre proximity from the sensor to heart surface, which enhances the cardiac magnetic field to greater than nanoteslas and allows the mapping of these signals with intra-cardiac resolution. From the acquired magnetic images, we also estimate the source electric current vector, flowing from the right atria base via the Purkinje fibre bundle to the left ventricular apex. Our results establish the solid-state quantum sensor's capability to probe cardiac magnetic signals from mammalian animals and reveal their intra-cardiac electrodynamics. This technique will enable the study of the origin and progression of myriad cardiac arrhythmias including flutter, fibrillation, and tachycardia.**

**Introduction**

Along with cancer and cerebrovascular disease, one of the leading causes of death globally is heart disease[1]. A root cause of many cases of heart failure, such as paroxysmal supraventricular tachycardia, atrial flutter, atrial fibrillation, and ventricular tachycardia, is the imperfection of electric current propagation occurring at the millimetre-scale[2–4]. In clinical settings, for example, the region of this imperfection is searched via electrophysiological study using multiple electrode catheters inserted under anaesthesia[5,6]. This search process, which often entails X-ray exposure and extends over several hours, is a primary pain point in the catheter ablation treatment for both patients and physicians. A non-invasive alternative approach is body surface electrocardiography (ECG)[7,8]. In ECG, however, the electric signal propagation needs to be inferred from the electric potential profile, which is modified considerably by the body's nonuniform and nonstationary electrical conductivity. These difficulties would be mitigated by mapping intra-cardiac currents more directly prior to surgery[9].

Magnetocardiography (MCG) is a contactless technique that remotely measures the stray magnetic fields produced by cardiac currents in the heart[10,11]. MCG meets implementation requirements using magnetic sensors such as superconducting quantum interference devices (SQUIDs)[12,13], optically pumped magnetometers (OPMs)[14–16], and tunnelling magnetic resistance (TMR) sensors[17,18]. Nevertheless, these sensors typically provide a resolution of only a few centimetres, which is fundamentally limited by either the standoff distance or sensor size. SQUIDs and OPMs operate with standoff distances of a centimetre or more from the biological target sample because their sensor head needs to be under severe temperature conditions (cryogenic and heated, respectively) and they have extended sensor geometries[13,19,20]. On the other hand, TMR offers an insufficient sensitivity of ~1 nT·Hz$^{-1/2}$ when it is scaled down to a millimetre size[17,18]. Therefore, the next milestone of MCG towards it being of more practical use than electrophysiology is to improve its spatial resolution to millimetre-scale by reducing the sensor volume and standoff distance while maintaining the sensitivity at a high enough level to detect those cardiac signals.

A quantum sensor based on an ensemble of nitrogen-vacancy (NV) centres in diamond[21–23] provides a volume-normalized magnetic field sensitivity of ~1 pT·Hz$^{-1/2}$·mm$^{3/2}$ and can be operated at room temperature[24,25]. Because of these advantages, combined with its ease of construction, the NV-based quantum sensor has been applied to various biological studies. Recent demonstrations include the magnetic imaging of living cells at the sub-cellular scale[26–28], magnetic detection of action potentials in living giant axons[24] and muscle fibres[29], and observation of biomagnetic nanoparticles for cancer diagnosis[30]. In this work, we report the first application of this quantum sensor to the measurement of magneto-cardiac signals generated from a live rat specimen with a millimetre-scale spatial resolution. This demonstration is realized by several technical advancements: a sensor structure that allows millimetre proximity to the heart without causing damage, a decoupling scheme of temperature fluctuations due to biological activity, and two cardiac current estimation methods that complement each other to reveal the spatiotemporal dynamics of the current. Below, we describe our experimental findings in detail and discuss them in light of novel cardiac current source models.

**Results**

Figure 1a illustrates our experimental setup. The rat specimens used in this study were 10-11-week-old males, anaesthetized, thoracotomized, and maintained for 5 h using an artificial respirator. The core of our sensor consisted of a single-crystal diamond chip containing high-density (~$8 \times 10^{16}$ cm$^{-3}$) electronic spins associated with negatively charged NV centres oriented in the z-direction of the laboratory frame. The ground-state energies of the NV centre $m_s = \pm 1$, which depend on an external magnetic field, were interrogated with a green laser and microwaves at room temperature (Fig. 1b). An ensemble of ~$1.5 \times 10^{13}$ of these NV centres in a laser illumination volume of 0.19 mm$^3$ was used to detect the z-component of the magnetic fields generated by electric currents flowing through each rat's heart (with a thickness of ~11 mm) placed directly under the sensor with 0.6-2.0 mm proximity relative to the heart surface (Fig. 1c). As explained in Fig. 1d, the time-varying cardiac magnetic field $B_z(t)$ was converted to a change in fluorescence from the NV centres using the frequency-modulated optically detected magnetic resonance (ODMR) scheme[24,29,31].

As a benchmark, we first evaluated the magnetic field sensitivity, stability, temporal resolution, and field dynamic range of our sensor in the absence of a rat. Figure 1e presents a measured magnetic field sensitivity $\eta_m = 140$ pT·Hz$^{-1/2}$ across the rat's cardiac signal bandwidth of ~200 Hz, determined from the sensor's noise floor. This magnetic field sensitivity, 7 times above the shot noise ($\eta_\gamma = 19$ pT·Hz$^{-1/2}$), was achieved using a threefold noise suppression technique: laser fluctuation cancellation by subtracting a pick-off laser beam signal from the red fluorescence signal[24,29], low-frequency electronic noise avoidance by lock-in up-converting[24,29] at a typical modulation frequency of $f_{mod} = 17$-25 kHz, and temperature drift compensation by monitoring double resonance peaks corresponding to the $m_s = 0 \leftrightarrow \pm 1$ transitions simultaneously[32]. This level of sensitivity was maintained stably over 3 h using a microwave feedback system, which compensated the long-term resonance frequency drift caused by environmental magnetic and thermal noise. The temporal resolution, estimated to be $\Delta t = 3.4 \times 10^2$ μs from the 10%-90% rise time with a roll-off of 24 dB/octave and a lock-in time constant of $\tau_{LIA} = 50$ μs, was fast enough to capture the signal profile. With a full-width-half-maximum ODMR linewidth of 2.1 MHz and a lock-in deviation frequency of $f_{dev} = 360$-400 kHz, the approximate magnetic field dynamic range, with a less than 3% loss in sensitivity, was estimated to be ±2.5 μT, comfortably above the typical magneto-cardiac signal of nanoteslas. Thus, we concluded that our sensor offered a sufficiently high performance for detecting the rat's cardiac magnetic fields.

Next, we performed MCG measurements concurrently with ECG signal detection. The obtained MCG signal was assessed from the following five viewpoints. First, as presented in Fig. 2a, the MCG signal and corresponding ECG data of rat A exhibited the same periods of the cardiac pulse cycle, manifesting that the magnetic signals originated from the rat's cardiac activities. Second, the MCG signals of two additional rats (labelled B&C) were acquired, supporting the reproducibility among rats A-C (Fig. 2a-c). Third, our MCG signal presented a similar profile to what was obtained from rat D with an OPM (Fig. 2d), fulfilling inferential reproducibility between the two different methodologies. Fourth, both the NV and OPM data were explained by the same electric current dipole model (see Methods) with the standoff distance from the centre of the heart to the sensor as a predictor variable (Fig. 2e). This agreement encourages that these sensors can also be used complementarily for extending the measurement space coverage. Fifth, the signal-to-noise ratio (SNR), defined as the mean signal peak amplitude ratio to the root mean square noise, scaled as the square root of the number of averages, indicating that the obtained MCG signal was a consequence of averaging over consistent repetitive peaks rather than random sporadic peaks (Fig. 2f). Furthermore, we excluded various other spurious signal sources, such as physical vibration of the diamond, the contamination of light reflected from the rat, and diamond temperature fluctuations due to the rat's body heat, that were synchronized with the rat's cardiac activity. From these considerations, we concluded that the acquired signal was the rat's cardiac magnetic field.

As shown in Figs. 3 and 4, the solid-state quantum sensor has the ability to perform rapid cardiac magnetic field mapping, revealing intra-cardiac current dynamics. In the following measurements, the diamond was fixed to the laboratory frame, while rat E was scanned horizontally in two dimensions with respect to the diamond across 11 × 11 pixels with a step size of 1.5 mm, covering most of the heart within a field of view of 15 mm × 15 mm (Fig. 3a). The measured standoff distance between the sensor and heart centre was $d^{NV} = 7.5(5)$ mm. Under appropriate imaging conditions, the magnetic field patterns produced by the rat could be measured within 40 seconds per pixel plus 40-minute preparation time with minimal tissue optical and microwave radiation exposure, such that no visible damage was observed after magnetic imaging and ECG recording for ~2 h. The obtained time-series magnetic data at each pixel were sliced at various timings, and corresponding magnetic images were constructed. In particular, the magnetic image at the R-wave peak presented a dipolar pattern with a pole-to-pole spacing of $\Delta^{NV} = 9.7(6)$ mm (Fig. 3b). This spreading was smaller than that obtained from rat D with an OPM at a measured standoff distance of $d^{OPM} = 14(2)$ mm (Fig. 3c).

To further characterize the properties of the source electric current, we fitted each of the two measured magnetic field patterns to a magnetic field pattern simulated with a multiple electric current dipole model. The obtained total dipole moment for the NV data, $Q^{NV} = 1.3(5) \times 10^3$ µA·mm, overall agreed with that obtained with the OPM, $Q^{OPM} = 1.0(4) \times 10^3$ µA·mm. By correlating with magnetic resonance imaging (MRI) (Fig. 3d), we also found that the central current dipole, which is the geometric centre of the current flow in the heart, was located within the left atrium towards the Purkinje fibre bundle (Fig. 3e and f). In this measurement, the resolution of the dipole moment $\Delta r_Q$ derived under Sparrow's criterion was 5.1 mm for the NV and 15 mm for the OPM, manifesting the intra-cardiac scale resolving power of the solid-state quantum sensor.

We highlight that our magnetic field imaging results can be used to determine the time-varying source electric current density distribution $\vec{J}(\vec{r})$ on a two-dimensional plane, which can reveal spatiotemporal information about the cardiac electrodynamics. The current density was calculated using bfieldtools[33,34], a software package that employs a stream-function method to rapidly reconstruct the magnetic potential parameters via nonlinear optimization. For simplicity, the current was approximated to flow within a flat two-dimensional plane with a standoff distance $d_Q^{NV} = 8.1(7)$ mm, as determined from the electric current dipole fitting. For spatial information, Fig. 4a and b present the measured magnetic field pattern (same as Fig. 3b) and estimated current density distribution sliced at the R-wave peak, revealing a current stretched in the left atrium near the Purkinje fibre. The total current flowing through the heart, calculated by integrating the density across a linecut (Fig. 4c) between the magnetic poles, was $I_J^{NV} = 2.0(3) \times 10^2$ μA. For temporal information, Fig. 4d-f presents the measured magnetic field pattern, calculated current density distribution, and current density across the line cut, evaluated at 20 ms after the R-wave peak. As expected, almost no signal was present, resulting in a total current of $I_J^{NV} = 13(3)$ μA. This result suggests that this current density estimation method can determine the distribution of current at various timings of the electric signal propagation.

**Discussion**

In summary, we exploit a key strength of the solid-state quantum sensor—the ability to bring the NV centres into proximity to the signal source under ambient operating conditions—to demonstrate the MCG of living mammalian animals and the associated electrical current estimation with a spatial resolution smaller than the heart's feature size. The two current estimation methods determine the geometric centre of the current in three dimensions and associated current density distribution on a projected two-dimensional plane as complementary information. The millimetre-scale MCG technique reported here is an essential step towards developing a rapid non-invasive diagnostic tool for various cardiac diseases. As an initial step towards this goal, our sensor combined with the two current models will allow more precise observation of the origin and progression of cardiac imperfections, such as spiral re-entry and abnormal automaticity involved in the pathogenesis of tachyarrhythmias[2,3], by using small mammalian model animals.

The applicability of our sensor can be extended further from MCG to various other biological current-driven phenomena via the following step-by-step technical improvements in magnetic field sensitivity. First, the sensitivity of our sensor approaches ~20 pT·Hz$^{-1/2}$ through optical engineering, which provides a severalfold improvement in the laser absorption rate and fluorescence collection efficiency. Successful approaches include multiple total internal reflections of the excitation laser beam[32] and fluorescence collection through a trapezoidal-cut diamond chip combined with a parabolic concentrator lens[35] or with directly attached photodiodes[36,37]. Second, ~1 pT·Hz$^{-1/2}$ can be achieved via incorporation of the pulsed NV control scheme. Promising approaches recently demonstrated include the double quantum 4-Ramsey protocol[38] and continuously excited Ramsey protocol[25]. Third, introducing additional quantum manipulation techniques such as magnetic impurity decoupling techniques[39] for extending the NV coherence time and quantum-assisted readout schemes[40,41] for enhancing the NV readout fidelity would yield a sensitivity of ~50 fT·Hz$^{-1/2}$. This sensitivity will enable detection of action potential propagation in muscle fibres, cerebral tissues, and spinal cord.

Finally, we also envision that our solid-state quantum sensor will add tremendous value to medicine and healthcare once the three-dimensional reconstruction of source currents at the millimetre scale becomes available by fully exploiting the following other unique advantages of NV magnetometry. Parallelization with CCD/CMOS camera-based detection[42,43] will boost the data acquisition speed linearly with respect to the number of image pixels. Full vectorization of magnetic sensing by encoding all four crystallographic classes of NV centres into separate lock-in modulation bands[31] will significantly facilitate the solving of the inverse problem of source reconstruction[44]. In addition, miniaturization and integration via on-chip fabrication[37] may allow the NV sensor to be installed on a catheter and an endoscope. This next-generation diamond sensor with these unique features and broad modalities, combined with ~50 fT·Hz$^{-1/2}$ sensitivity and ambient operating conditions, would meet numerous medical and healthcare applications, including rapid localization of the cause of acute cardiac fibrillation and epilepsy, non-invasive diagnosis of a new foetus's cardiac/cerebral functions, robot-assisted rehabilitation and augmentation of body functions, real-time monitoring of physical conditions for personalized medicine and accident prevention during automobile driving and industrial machine operation.

**Methods**

**Rat surgical protocol:** The animal experiment was approved by the University of Tokyo Ethical Review Board (reference number KA18-15), and every procedure followed the institutional guidelines, ensuring the humane treatment of animals. The animals studied were five male SLC/Wistar rats (Japan SLC, Inc., Tokyo, Japan) aged 10–11 weeks. Each rat was held on a hot-water bed (water temperature: 45.0 °C), where hot water was circulated through a silicon tube via a water circulation system (Thermo Haake DC10/K10, Thermo Fisher Scientific GmbH, Germany) for maintaining its body temperature. The anaesthesia, tracheotomy, artificial ventilation, and thoracotomy processes were as follows. First, rats were anaesthetized using 2–3% isoflurane in 300 mL/min air via an automatic delivery system (Isoflurane Vaporizer; SN-487; Shinano Manufacturing, Tokyo, Japan). Next, under moderate anaesthesia, the body hair of each rat was shaved, and tracheotomy was performed for artificial ventilation. For the artificial respirator (Small Animal Ventilator, SN-480-7, Natsume Seisakusho Co., Ltd., Japan), 2–3% isoflurane in 2.5 mL air per respiration at 80 times/min was delivered to each rat during the imaging of cardiac magnetic fields and the surgical process. Subsequently, thoracotomy was performed to expose the heart. Nylon threads were used to lift the heart for further exposure outside the body surface. After completion of all experimental procedures, the animals were sacrificed under deep anaesthesia.

**Magnetic resonance imaging:** Before magnetic field measurements on rat samples, MRI was conducted with a 7-Tesla MRI system (BioSpec 70/20USR, BRUKER, Germany) to confirm each heart's internal structure. Each rat was mounted in a cylindrical sample holder. Images of the rat hearts were obtained without a contrast agent and using a FLASH-cine sequence at 1-mm thickness and with a 60 mm × 60 mm field of view. The details of the MRI sequences are as follows: repetition time $T_R$ = 2.5 ms, echo time $T_E$ = 8.0 ms, 192 × 192 pixels, excitation pulse angle = 15°, exposure time for movie recording = 160 ms, and number of movie cycles = 20.

**Electrocardiography:** Each sample's electrode voltage was recorded with an ECG recording instrument (Neuropack X1, Nihon Kohden Corporation, Japan) through a recording electrode (Natus

Ultra Subdermal Needle Electrode 0.38 mm, Natus Neurology Inc., USA) before being sent to the data acquisition module simultaneously with the MCG signal. The ECG signal was used as a reference for the MCG measurement and to monitor heart activities[45]. The typical heart rate was 4.5-8.0 Hz.

**NV diamond sample:** The NV diamond used in this study was prepared by the following procedure. A single-crystal diamond was synthesized under high-pressure and high-temperature (HPHT) conditions. The carbon source had a natural abundance of isotopes (1.1% $^{13}$C). After diamond synthesis, the grown crystals were cut parallel to the {111} crystal plane, and both the top and bottom surfaces were mirror-polished. The obtained HPHT {111} crystal was a truncated hexagonal pyramid with approximate dimensions of 5.2 mm$^2$ × 0.35 mm. It was then irradiated with a 2.0 MeV electron beam at room temperature with a total fluence of $5 \times 10^{17}$ electrons·cm$^{-2}$, followed by annealing at 1,000 °C for 2 h under a vacuum. The concentrations of the P1 centres and NV$^-$ centres in this NV diamond sample were measured by CW electron spin resonance (ESR) to be 15 ppm ($2.6 \times 10^{18}$ cm$^{-3}$) and 1.8 ppm ($3.2 \times 10^{17}$ cm$^{-3}$), respectively. The relative uncertainties in these concentrations were both roughly ± 30%.

**NV physics:** The NV centre ground state is a spin triplet ($S = 1$) composed of $m_S = 0$ and $\pm 1$ Zeeman sublevels split by 2.87 GHz owing to spin−spin interactions. This zero-field splitting is susceptible to the diamond temperature[46], with a temperature–frequency coefficient of approximately −74 kHz·K$^{-1}$. An external magnetic field $B_0$ shifts the $m_S = \pm 1$ states in Hertz by $\pm g_e \mu_B h^{-1} B_0$, where $g_e = 2.003$ is the electron g-factor of the NV$^-$ ground state, $\mu_B = 9.274 \times 10^{-24}$ J·T$^{-1}$ is the Bohr magneton, and $h = 6.626 \times 10^{-34}$ J·s is the Planck constant. Because of the hyperfine interaction with the host $^{14}$N nuclear spin ($I = 1$) and its nuclear quadrupolar interaction, each sublevel further splits into three hyperfine states $m_I = 0, \pm 1$. Illumination by green (532 nm) laser light excites the NV centres from these ground states to excited states while conserving their spin. When the NV decays from the excited states, it fluoresces red (637–800 nm) light. A nonradiative, spin non-conserving intersystem crossing decay path from the $m_S = \pm 1$ excited states to the $m_S = 0$ ground state through metastable singlet states allows optical polarization of the ground state to $m_S = 0$. The amount of emitted red fluorescence can distinguish the $m_S = 0$ state from the $m_S = \pm 1$ states. In CW ODMR spectroscopy, the fluorescence intensity presents a dip when the microwave frequency is resonant with the NV transition frequency between the $m_S = 0$ and $\pm 1$ ground states as the microwave frequency is swept with continuous laser illumination.

**Solid-state quantum sensor structure:** Our sensor design is based on that of Schloss *et al.*[31] However, we sterically constructed the optical system on two stories to put the system in a custom-made magnetically shielded room with four layers of permalloy (Ishida Ironwork's Co., Ltd.). A 532-nm laser (Coherent Verdi-G5) was installed on the first floor of the setup. The laser polarization was adjusted to p-polarization with a half-wave plate (Thorlabs WPH05M-532). The laser beam was directed upward, guided through an M6 screw hole on a breadboard. On the second floor, the beam was then passed through a lens with a focal length $f$ = 400 mm (Thorlabs LA1172-A) and directed diagonally downward by a silver mirror (Thorlabs PF10-03-P01). A laser beam with a 1/e² Gaussian width of ~400 μm and a typical incident power $P_0 = 2.0$ W was introduced to the diamond's top major (111) surface at an incidence angle of ~70°. The estimated transmittance at the top air–diamond interface was > 99%. An achromat 1.25 NA Abbe condenser lens (Olympus U-AC) collected red fluorescence from the NV centres through the top surface. Fluorescence was then passed through a long-pass filter (Thorlabs FELH0600) and directed onto a silicon photodiode (Thorlabs SM1PD1A). The condenser lens, the optical filter, and the photodiode were mounted downward on an XY manual translation stage (Thorlabs XYT1/M). Their heights were adjusted using a long-travel vertical translation stage (Thorlabs VAP10/M). The typical power of the collected fluorescence was $P_F = 33$ mW, corresponding to a photocurrent of 14 mA, given the photodiode's 0.45 A/W responsivity at a 680 nm wavelength. A beam sampler (Thorlabs BSF-10A) was used to pick off ~1.5% of the laser light, which was directed onto another silicon photodiode (Thorlabs SM1PD1A). This photocurrent signal became a reference for cancelling the laser fluctuation noise. A rare-earth magnet (Magfine Corporation NR0101) aligned along the [111] orientation applied a uniform static bias field of $B_0 = 1.4$ mT at the diamond to split the $m_S = \pm 1$ ground states. The microwave was irradiated on the diamond via a homemade microwave antenna. Each rat was placed on a 33 cm × 23 cm × 2.1 cm custom-made acrylic board with a hot water circulating heater. The acrylic board was placed on a manual translation stage (Thorlabs L490/M) for height adjustment (Z-axis) and an automatic XY translation stage (SIGMA KOKI HPS120-60XY-SET) for horizontal mapping (X-Y plane).

**Lock-in DC magnetometry scheme:** In lock-in DC magnetometry, the microwave carrier frequency is set to a value at which the lock-in ODMR slope becomes maximal while the NV states are continuously excited via optical and microwave illumination. A time-varying external DC magnetic field $B(t)$ is then sensed as the shift in the resonance frequency $f_0(t) = f_0 + \delta f(t)$, where $\delta f(t) = g_e \mu_B h^{-1} \delta B(t)$. When the resonance frequency changes, the fluorescence intensity and the lock-in signal change accordingly. The field is then extracted by dividing the change in the lock-in voltage $\delta V$ by the lock-in ODMR slope $dV/df$ as $\delta B(t) = g_e^{-1} \mu_B^{-1} h \, \delta V(t)(dV/df)^{-1}$. We applied this frequency up-modulation to both $m_S = 0 \leftrightarrow \pm 1$ transitions simultaneously but with different modulation frequencies $f_{mod}^{\pm}$ and deviation frequencies $f_{dev}^{\pm}$; thus, information about each transition was encoded in different modulation frequency bands. The signal associated with each transition was obtained by lock-in demodulation at the corresponding modulation frequencies. This method allowed us to cancel out the temperature fluctuations that appeared as common-mode noise in both transitions.

**Signal processing:** In this experiment, we obtained four lock-in amplifier outputs: $V_F^+, V_F^-, V_L^+, V_L^-$, where $V_F^{\pm}$ is the fluorescence signal demodulated at $f_{mod}^{\pm}$ and $V_L^{\pm}$ is the laser reference demodulated at $f_{mod}^{\pm}$. The data analysis process comprised five steps. First, laser noise cancellation was performed. Laser fluctuations appeared in all four outputs as common-mode noise. This noise can be suppressed by subtraction: $V_s^+ = V_F^+ - \xi^+ V_L^+$ and $V_s^- = V_F^- - \xi^- V_L^-$. The scaling factors $\xi^{\pm}$ were fixed to the ratio between the fluorescence photodiode's average signal level and the laser reference photodiode's average signal level, measured at the beginning of the measurement run. Second, the temperature drift that interferes with cardiac magnetic field detection was cancelled as common-mode noise by subtracting the $m_S = 0 \leftrightarrow +1$ signal from the $m_S = 0 \leftrightarrow -1$ signal: $V_s = (V_s^- - \zeta V_s^+)/2$. The scaling factor $\zeta$ was fixed to the ratio between the $m_S = 0 \leftrightarrow \pm 1$ signal slopes, determined from the lock-in ODMR spectrum. Third, notch filtering was performed. The electronic background noise within the measurement bandwidth was removed with a notch filter at multiples of 11 and 50 Hz. The filtered signal was defined as $V_s'$. Fourth, we applied signal timing compensation. Because the cardiac signal frequency drifts over time due to the rat physiology, the magnetic pulse timing requires adjustment using the ECG pulse peak location as a timing reference. Fifth, band-pass filtering was performed. Because most of the magneto-cardiac signals were between 3 and 200 Hz, a 3–200 Hz band-pass filter was applied.

**OPM magnetocardiography:** To verify the experimental data measured by the solid-state quantum sensor, we performed two-dimensional MCG imaging with an OPM (QZFM-Gen2, QuSpin Inc.) inside the magnetically shielded room. The OPM sensor's vapour cell was located 8.5 mm above each rat's heart surface. A two-dimensional image of the MCG signals was recorded on the X–Y plane ([−15, +15] mm with a 3 mm step size, 11 × 11 = 121 measurement points, and a 1 kHz sampling rate).

**Electric current dipole model**: The multiple electric current dipole model used in this work consisted of $N_Q = 7$ central current dipoles with the same magnitude and orientation and a pair of return current dipoles with opposite orientations. The central dipoles were distributed evenly across the vertical cross-section of the heart: $\vec{Q}_C = \sum_{i=1}^{N_Q} \vec{Q}(\vec{r}_0 + \vec{z}_i)/N_Q$, where $\vec{Q}$ is the total electric current dipole moment, $\vec{r}_0$ is the position vector of the geometric centre of the dipoles, and $\vec{z}_i = (0, 0, z_i)$ is the vector of the location of each dipole from the centre. The return current dipoles were placed $\pm\vec{\rho}_R = \pm(x_R, y_R, 0)$ from the centre: $\vec{Q}_R = -k_R[\vec{Q}(\vec{r}_0 + \vec{\rho}_R) + \vec{Q}(\vec{r}_0 - \vec{\rho}_R)]$, where $k_R$ is the ratio of the return current amplitude and the vectors $\vec{Q}$ and $\vec{\rho}_R$ are set orthogonal to each other. The fitted curve for Fig. 2e was calculated by nonlinear least squares regression with a fitting parameter of $Q = |\vec{Q}|$, and a predictor variable of the standoff distance $d$. The dipoles in Fig. 3 were calculated by matching simulated magnetic field images with those measured using an L$^2$-norm minimization routine. The fitting parameters were the magnetic moment vector $\vec{Q} = (Q_x, Q_y, 0)$, the magnetic moment centre location $\vec{r}_0 = (x_0, y_0, d_Q)$, and the return current dipole location $\vec{\rho}_R$. Here $d_Q$ is the standoff distance between the centre of the central dipoles and the sensor. MRI measurements of the relative sizes and positions of each rat's heart were used to provide an initial guess and impose constraints on these fitting parameters. The uncertainty of the fitted parameters $\delta Q_x, \delta Q_y, \delta x_0, \delta y_0, \delta d_Q$ was estimated from the 68% confidence interval of the fitting. The obtained standoff distance $d_Q \pm \delta d_Q$ was then used in the current density estimation presented in Fig. 4.

**Electric current density model:** The process of electric current density estimation from the obtained magnetic field images in Fig. 4 employed bfieldtools[33,34], an open-source Python software suite. In this software, the surface-current density $\vec{j}(\vec{r})$ originates from a piecewise linear stream function $\psi(\vec{r})$. This stream function is determined such that the L²-norm of the difference between the simulated field $B_{sim}$, which is derived from the stream function using the Biot-Savart's law, and the measured field $B_{meas}$ under a penalty term with a strength of $\lambda$ becomes minimal: $\psi \in \mathrm{argmin}\, \|B_{meas} - B_{sim}(\psi)\|^2 + \lambda \|\psi\|^2$. Vectors along the contour lines of the stream function represent the surface current density $\vec{j}(\vec{r}) = \nabla_\| \psi(\vec{r}) \times \hat{n}(\vec{r})$, where $\hat{n}$ is the unit vector perpendicular to the surface and $\nabla_\| = \nabla - \hat{n}(\hat{n} \cdot \nabla)$ is an operator that projects the gradient vector of a scalar function onto the tangent plane of the surface.

**Spatial resolution**: In this work, we evaluated the resolution and localization precision of the electric current dipole moment and magnetic field. For the electric current dipole moment $\vec{Q}$, the resolution $\Delta r_Q$ (5.1 mm for the NV and 15 mm for the OPM) was calculated by applying Sparrow's criterion to the associated magnetic field pattern. The localization precision $\delta r_Q$ (1.2 mm for the NV and 2.9 mm for the OPM) was defined as the uncertainty (one standard deviation) of the fitted current dipole moment position. For the magnetic field image $B_z(x, y)$, the resolution $\Delta r_B$ (1.5 mm for the NV and 3.0 mm for the OPM) was limited by the imaging pixel size. The localization precision $\delta r_B$ (0.4 mm for the NV and 0.4 mm for the OPM) was determined from the uncertainty (one standard deviation) of the fitted magnetic pole position.

**Acknowledgements**

We would like to thank Hiromitsu Kato for helpful discussions about the diamond preparation, Ichiro Sakuma for the valuable advice on the potential applications of solid-state quantum sensors to the magnetocardiography of other animals, and Hitoshi Ishiwata, Kosuke Mizuno, and Takeharu Sekiguchi for their comments regarding the manuscript. This work is supported by the MEXT Quantum Leap Flagship Program (MEXT Q-LEAP), Grant Numbers JPMXS0118067395 and JPMXS0118068379. KA receives funding from the JST, PRESTO, Grant Number JPMJPR20B1.

**Author contributions**

The project was conceived by KA and AK. The magnetometer was designed and constructed by KA, DN, YN, and YH. The rat dissection and anaesthesia were conducted by AK and ZX. The MCG/ECG data collection and analysis were conducted by KA, AK, DN, RM, ZX, YN, and XC. KA, AK, DN, IF, RM, and XC developed the cardiac signal model and performed electric current reconstruction. AK and ZX acquired the OPM and MRI images. The NV diamond was grown by CS, MM, T Taniguchi, and T Teraji and was electron-irradiated by SO and TO. MY advised the interpretation of ECG/MCG results. This manuscript was prepared by KA and AK with review contributions from all the other authors. The overall supervision was performed by MH, MS, and TI.

**Competing interests**

The authors declare that they have no known competing interests that would influence the work reported here.

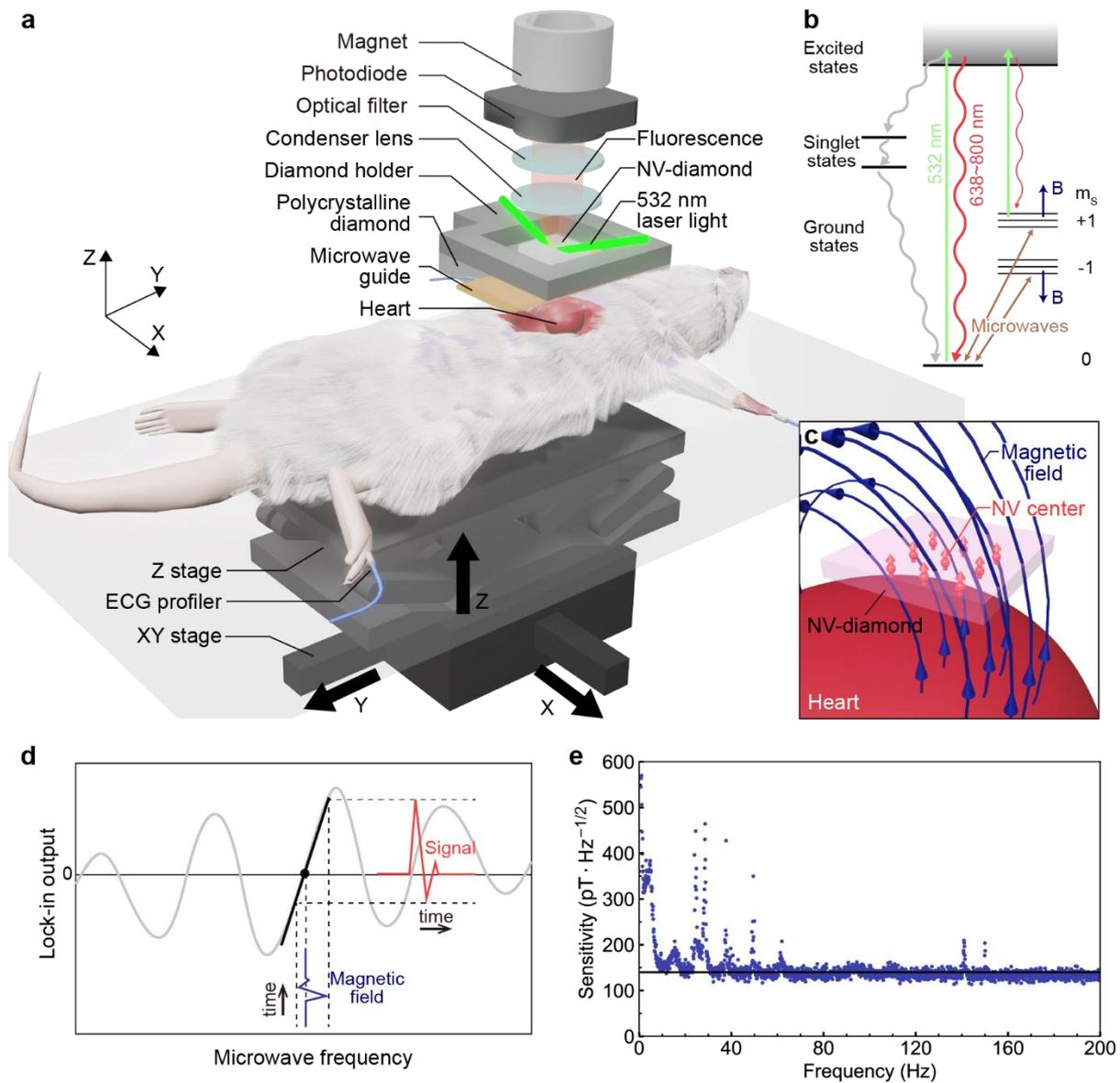

**Figure 1: Magnetocardiography based on a solid-state quantum sensor**. **a**, Schematic of the rat MCG setup. A living rat's heart remains approximately one millimetre below a diamond chip containing an ensemble of NV centres. The rat is scanned automatically along the XY-axes for magnetic field mapping and manually along the Z-axis for height adjustment. An electrocardiography (ECG) signal is monitored through ECG profilers concurrently with the MCG. The NV centres are excited by a 2.0 W green laser light introduced at an incident angle of ~70°. This excitation entails spin-state-dependent fluorescence collected by an aspheric condenser lens. **b,** NV centre energy level diagram. The $m_S = \pm 1$ ground states are split by a uniform 1.4 mT bias field applied by a permanent magnet and mixed by microwaves resonant with the NV transition frequency. Each of the ground states are further split by hyperfine interactions with the host $^{14}$N nuclear spin. **c**, Enlarged view of the heart and diamond. Electric currents (not shown) flowing through the heart generate a circulating magnetic field (blue arrows). The NV centres (red arrows) along the [111] orientation are sensitive to the Z-component of the magnetic field. **d**, Magnetometry principle. The time-varying target cardiac magnetic field (blue), which shifts the NV transition frequency, is converted to a change in the lock-in-demodulated fluorescence signal (red). Five peaks are observed in the lock-in ODMR spectrum because three hyperfine transition frequencies are excited with three-tone microwaves. **e**, Magnetic field sensitivity across the rat's heart signal frequency band of DC~200 Hz. As a guide for the eye, the black dashed line indicates 140 pT·Hz$^{-1/2}$.

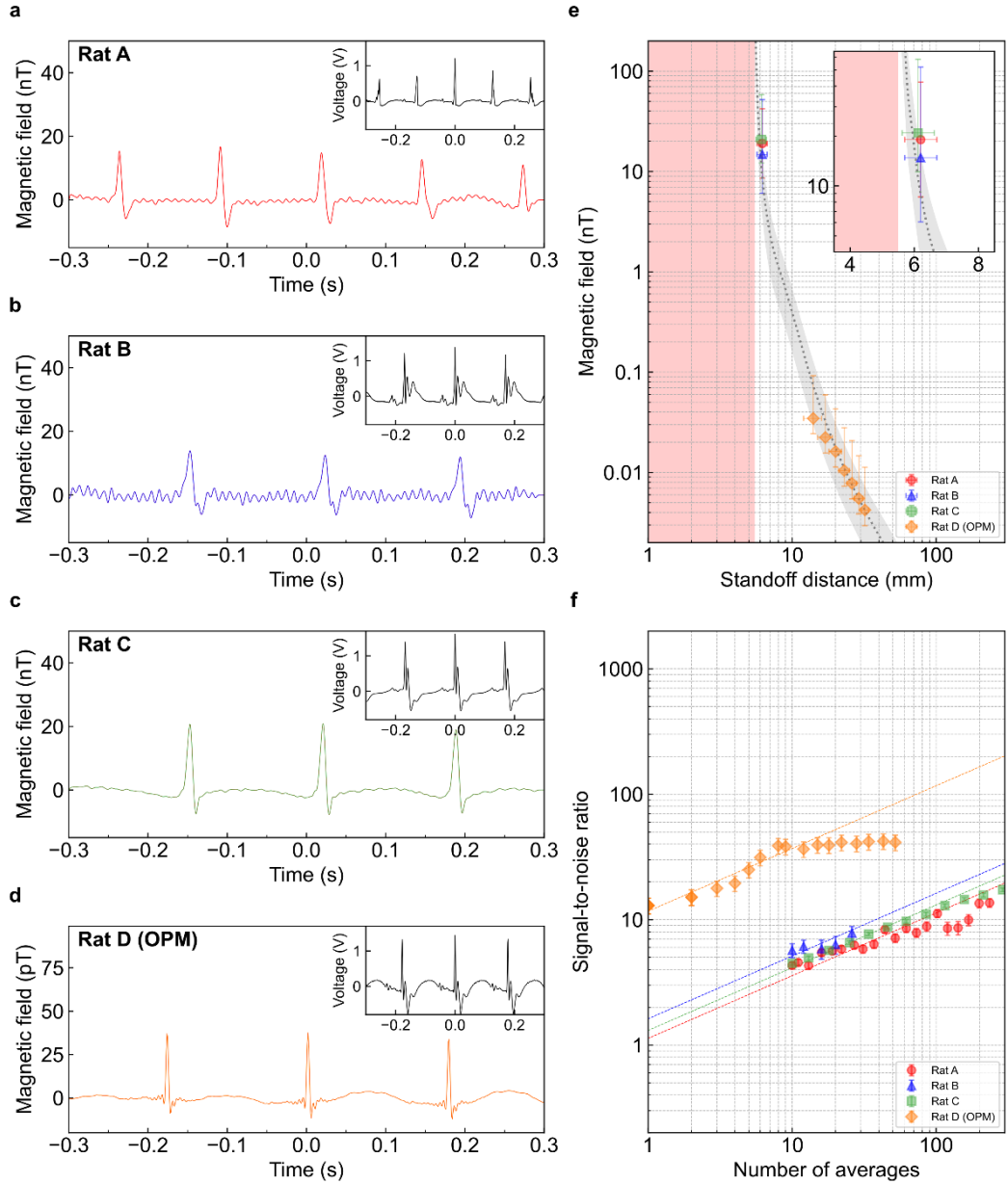

**Figure 2: Time-domain cardiac magnetic field signal**. **a-d**, Typical time-domain magneto-cardiac signal from rats A, B, C, and D, detected with the solid-state quantum sensor (rats A-C) and an OPM (rat D). The insets show ECG signals recorded simultaneously. Laser noise cancellation, temperature compensation, notch filtering at multiples of 11 and 50 Hz, signal averaging using the ECG signal as a timing reference, and band-pass filtering between 3 and 200 Hz were performed. The peaks correspond to the R-wave, the repetition rate of which matches that of ECG. **e**, Dependence of the cardiac magnetic field strength at the R-wave peak on the standoff distance $d$ from the heart centre to the sensor. The vertical error bars reflect an indeterminacy of the sensor position in the XY direction relative to the point of the maximal magnetic field. The horizontal error bars represent the systematic error due to an inaccuracy in measuring the standoff distance. The grey dashed line is the magnetic field calculated from a multiple electric current dipole model with a total dipole $Q = 1.4(2) \times 10^3$ μA·mm (fitted). The grey shaded area also reflects an indeterminacy of the sensor position. The pink shaded area represents the heart domain. **f**, Dependence of the measurement SNR on the number of averages of the cardiac pulse $N_{ave}$. Dashed lines show the fitted model $SNR \propto N_{ave}^{1/2}$. For the solid-state quantum sensor (rats A-C), square-root dependence is observed, while for the OPM, the SNR saturates at ~40 due to a long-term drift of the residual environmental magnetic field. Error bars are calculated from the signal, measurement noise, and the number of averages.

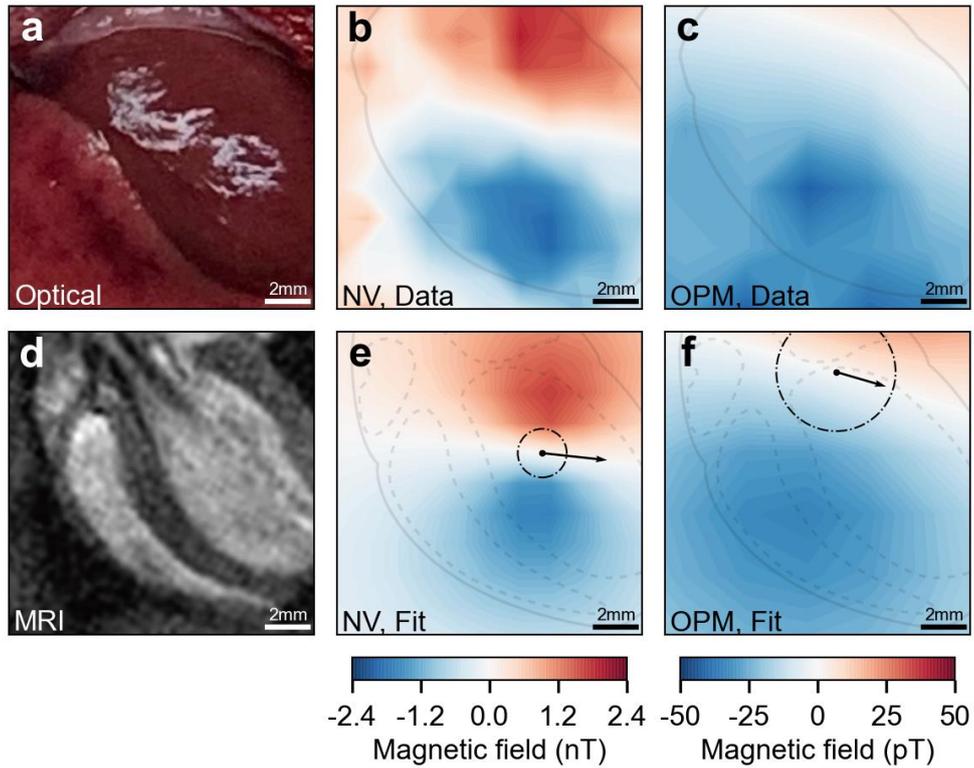

**Figure 3: Millimetre-scale cardiac magnetic field mapping. a**, Optical image of the rat's heart. **b-c**, Measured magnetic field map at the timing of the R-wave peak obtained with the NV centres for $d^{NV} = 7.5(5)$ mm and with the OPM for $d^{OPM} = 14(2)$ mm, respectively. The field of view is the same as that of the optical image with an accuracy better than 2 mm. The superimposed grey solid line shows the contour of the rat's heart traced from the optical image (**a**). The positive and negative poles are localized within the heart for the NV. **d**, Magnetic resonance image of the rat's heart, the orientation of which is adjusted to the optical image. **e**-**f**, Fitted magnetic field map using the multiple electric current dipoles model, superimposed on the contour of the internal structure of the rat's heart (grey dashed line) revealed by MRI (**d**). The black arrows represent the location, direction, and relative magnitude of the central dipole. The black dash-dotted circle depicts the localization precision of the central dipole. The electric current dipoles are estimated such that the mean squared error between the measured and simulated magnetic fields is minimal. The obtained central dipole ($Q^{NV} = 1.3(5) \times 10^3$ μA·mm for the NV and $Q^{OPM} = 1.0(4) \times 10^3$ μA·mm for the OPM) points to the lower right and is located in the left atrium of the rat's heart, indicating that the overall current flows from the right ventricles via the Purkinje fibre to the left atrium base. The resolution of central dipole $\Delta r_Q$ is estimated under Sparrow's criterion to be 5.1 mm for the NV and 15 mm for the OPM.

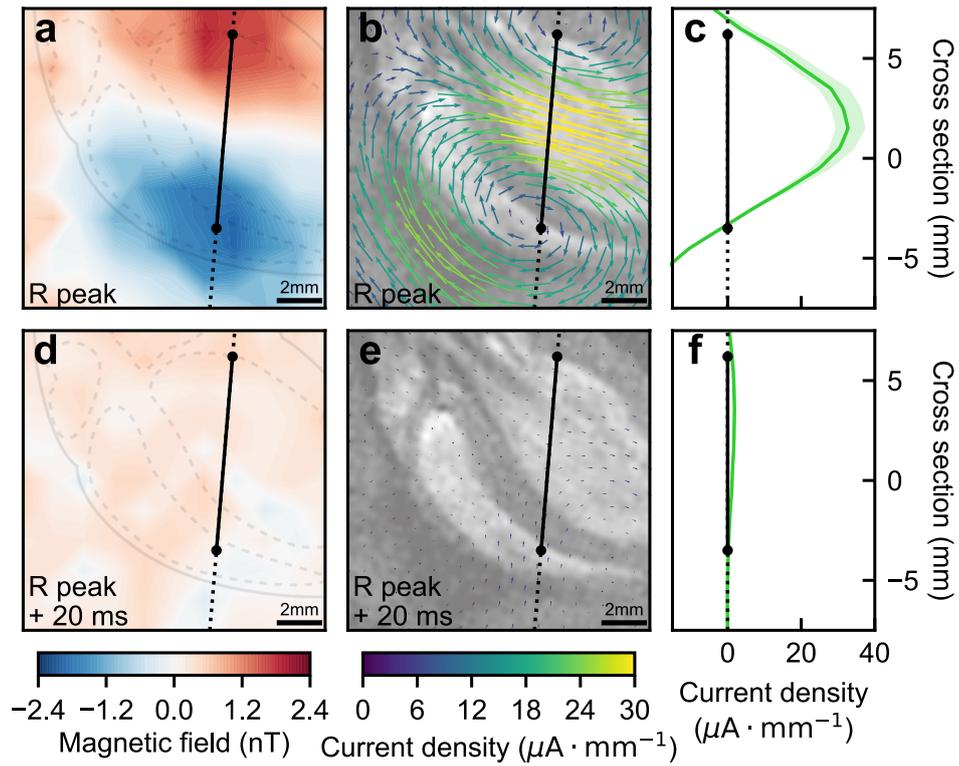

**Figure 4: Estimation of the spatiotemporal dynamics of the cardiac current**. **a,** Measured field map at the R-wave peak, superimposed on the contour of the structure of the rat's heart revealed by optical imaging and MRI. The black solid line is a linecut connecting the centres of the observed positive and negative magnetic field peaks. **b,** Vector plots of the electric current density calculated using bfieldtools with a standoff distance of $d_Q^{NV} = 8.1(7)$ mm. As a guide for an eye, the MRI image is superimposed in greyscale. **c,** Normal component of the electric current density vector with respect to the linecut. The green shaded region shows the uncertainty of the current density estimated from the uncertainty of the standoff distance (one standard deviation) obtained in the dipole fitting. Integrating the current density between the two dipole peaks yields the total current $I_J^{NV} = 2.0(3) \times 10^2$ µA crossing the linecut. **d-f,** Measured field map (**d**), estimated electric current density (**e**), and electric current density across the linecut (**f**) evaluated at 20 ms after the R-wave peak. The total current crossing the linecut is $I_J^{NV} = 13(3)$ µA.